\documentclass{PoS}

\usepackage{amsfonts} 
\usepackage{amssymb} 
\usepackage{amsmath} 
\usepackage{subfigure}
\usepackage{graphicx} 
\usepackage{array} 
\usepackage{dcolumn} 
\usepackage{bm} 
\usepackage{latexsym} 
\usepackage{longtable} 
\usepackage{color}
\usepackage{mathrsfs}
\usepackage{bbold}
\usepackage{multirow}
\usepackage{arydshln} 

\graphicspath{{./figs/}}
\newcommand{\fm}{\mathop{\rm fm}\nolimits}
\newcommand{\MeV}{\mathop{\rm MeV}\nolimits}
\newcommand{\GeV}{\mathop{\rm GeV}\nolimits}

\newcommand{\Tr}{\textrm{Tr}}

\title{ 
  Disconnected Quark Loop Contributions to Nucleon Structure
}

\ShortTitle{
  Disconnected Contributions to Nucleon Structure
}

\author{Tanmoy Bhattacharya, Rajan Gupta, \speaker{Boram Yoon} \\
  Los Alamos National Laboratory, MS B283, P.O. Box 1663, Los Alamos, NM 87545, USA \\
  E-mail: \email{boram@lanl.gov}}

\abstract{
We calculate the disconnected contribution to isoscalar nucleon charges for
scalar, axial and tensor channels of light and strange quarks.
The calculation has been done with the Clover valence quarks on the MILC 
$N_f=2+1+1$ HISQ lattices whose light quark masses corresponding to the pion 
masses of $305\MeV$ and $217\MeV$ at $a\approx0.12\fm$ and $312\MeV$ at 
$a\approx0.09\fm$.
All-mode-averaging technique is used for the evaluation two-point functions.
Disconnected quark loops are estimated by using the truncated solver method with
Gaussian random noise sources. 
Contamination from the excited states is removed by fitting the results of 
various source-sink separations and operator insertions to the formula including 
up to the first excited state, simultaneously.
}

\FullConference{The 32nd International Symposium on Lattice Field Theory,\\
    23-28 June, 2014\\
    Columbia University New York, NY}

\begin{document}

\section{Introduction} 
Disconnected quark line diagrams have non-trivial contribution to the lattice QCD 
study of nucleon structure.
However, direct calculation of the disconnected diagrams is not practical as they 
require all-to-all propagators.
Development of stochastic estimation for the disconnected diagrams made it 
feasible with current computing power, but it is still one of the most expensive
computations in lattice QCD \cite{Bali:2009hu}.

The simplest observable are the local bilinear operators that give the nucleon
charges $g_{A,S,T}$, nucleon $\sigma$ term and strangeness.
Those play an important role in dark matter search and study of neutron electric
dipole moment (nEDM) \cite{Bhattacharya:2013ehc}.
In this paper we report a lattice QCD calculation of the disconnected quark loop
contribution to isoscalar nucleon charges of light and strange quarks.
The calculation is done with clover valence quarks on the $N_f=2+1+1$ highly 
improved staggered quarks (HISQ) ensembles generated by MILC collaboration 
\cite{Bazavov:2012xda}.
The three ensembles that we use are listed in Table~\ref{tab:ens}.

\begin{table}[hbtp]
\begin{center}
\begin{tabular}{c|cccc}
\hline\hline
ID & $a$ (fm) & $M_\pi$ (MeV) & $L^3\times T$    & $M_\pi L$
\\ \hline
a12m310 & 0.1207(11) & 305.3(4) & $24^3\times 64$ & 4.54 \\
a12m220 & 0.1184(10) & 216.9(2) & $32^3\times 64$ & 4.29 \\
a09m310 & 0.0888(08) & 312.7(6) & $32^3\times 96$ & 4.50 \\
\hline\hline
\end{tabular}
\caption{ MILC HISQ lattices analyzed in this study.
Lattice parameters are obtained from Ref.~\cite{Bazavov:2012xda}.
}
\label{tab:ens}
\end{center}
\end{table}
%

\begin{figure*}[tb]
\centering{
  \subfigure{
    \includegraphics[width=0.35\linewidth]{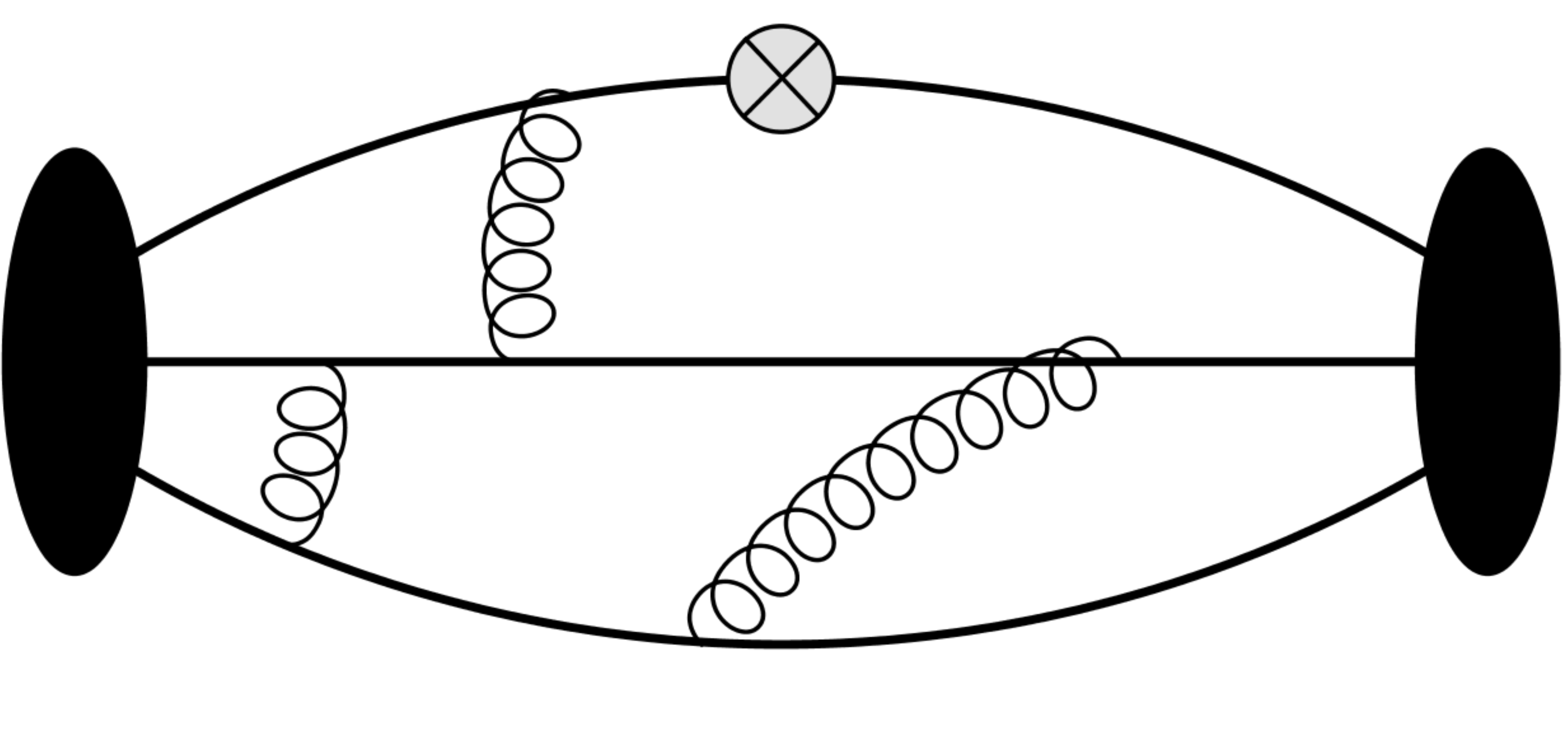}
  }
  \hspace{0.08\linewidth}
  \subfigure{
    \includegraphics[width=0.35\linewidth]{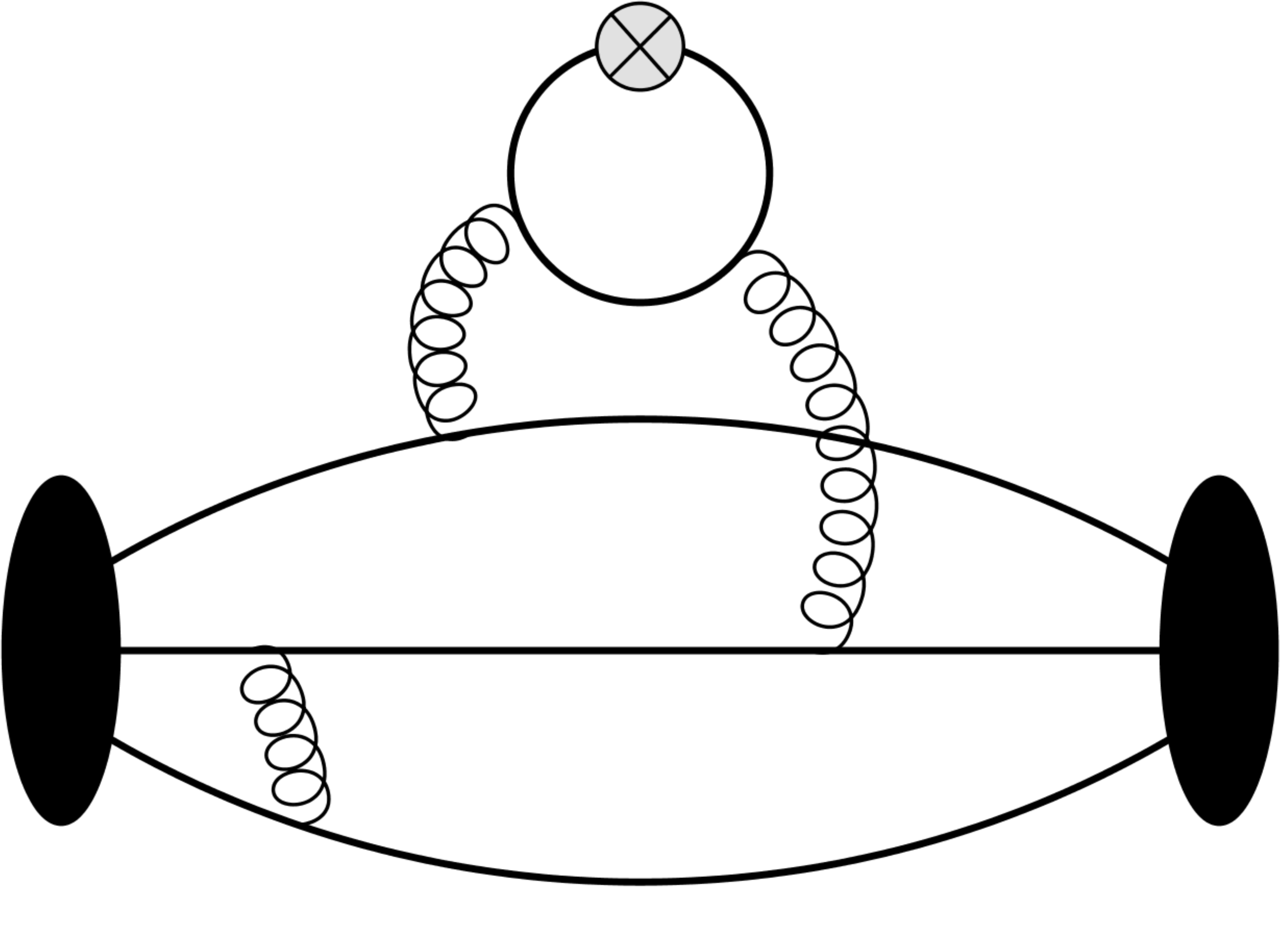}
  }
\caption{ Quark line connected (left) and disconnected (right) diagrams of the
  nucleon three-point function.
  \label{fig:con_disc}}
}
\end{figure*}

Nucleon charge $g_\Gamma^q$ of quark flavor $q$ and gamma structure $\Gamma$ can
be extracted from the ratio of the three-point function and the two-point function 
of nucleons as follows:
\begin{align}
 R_\Gamma(t, \tau) 
   \equiv \frac{\langle \Tr [ \mathcal{P}_\Gamma C_{\Gamma}^{\text{3pt}}(t, \tau) ]\rangle}
         {\langle \Tr [ \mathcal{P}_\text{2pt} C^{\text{2pt}}(t) ] \rangle}
   \xrightarrow{~ t \gg \tau \gg 0 ~} g_\Gamma^q\,.
 \label{eq:3pt_2pt_ratio}
\end{align}
%
%
Here $\Gamma$ is the gamma structure of the local bilinear operator in the three 
point function, $t$ is the source-sink separation, $\tau$ is the distance in 
Euclidean time between the source and inserted operator, and $\mathcal{P}_\Gamma$ 
and $\mathcal{P}_\text{2pt}$ are the projection operators.
There are two possible classes of contractions when we insert a bilinear operator 
between two nucleon states. One is to contract the bilinear operator with one of 
the three quarks of nucleons, as shown in the left hand side (l.h.s) diagram of 
Fig.~\ref{fig:con_disc}, and the other is to contract the bilinear operator by
themselves making a quark loop, as shown in the right hand side (r.h.s) diagram of
Fig.~\ref{fig:con_disc}.
%
%
%
Here we report our recent calculation of the disconnected quark loop contributions 
to the nucleon charges with various improvement techniques.

%

\section{Disconnected Quark Loop Contribution}
\label{sec:disc}
The disconnected quark loop part of Eq.~\eqref{eq:3pt_2pt_ratio} is written as
\begin{align}
 R_\Gamma^\text{dis}(t, \tau)
   =  \langle \sum_\mathbf{x} \Tr [M^{-1}(\tau, \mathbf{x}; \tau, \mathbf{x}) \Gamma] 
     \rangle 
  - \frac{\langle \Tr [ \mathcal{P}_\Gamma C^{\text{2pt}}(t) ] 
    ~\sum_\mathbf{x} \Tr [M^{-1}(\tau, \mathbf{x}; \tau, \mathbf{x}) \Gamma] 
    \rangle}
   {\langle \Tr [ \mathcal{P}_\text{2pt} C^{\text{2pt}}(t) ] \rangle}
 \,,
 \label{eq:3pt_2pt_ratio_disc}
\end{align}
with the Dirac operator $M$.
The calculation needs the evaluation of (1) nucleon two-point functions 
$C^{\text{2pt}}$ and (2) quark loops $\Tr[M^{-1}\Gamma]$.
%

\subsection{Two-point Function}
Exploiting translation symmetry of lattice, we average the nucleon two-point
function over multiple source positions.
In order to reduce the computational cost for the multiple source positions, we 
apply the all-mode-averaging (AMA) technique \cite{Blum:2012uh}.
Using AMA, the two-point functions for most of the source positions are 
calculated by a low-precision (LP) Dirac inverter with cheap computational cost.
The systematic error of the LP estimate is corrected by a few high-precision (HP) 
calculations of those as follows:
\begin{align}
 C^\text{2pt, imp}(t, t_0) 
 = \frac{1}{N_\text{LP}} \sum_{i=1}^{N_\text{LP}} 
    C_\text{LP}^\text{2pt}(t; t_0, \mathbf{x}_i^\text{LP})
  + \frac{1}{N_\text{HP}} \sum_{i=1}^{N_\text{HP}} \left[
    C_\text{HP}^\text{2pt}(t; t_0, \mathbf{x}_i^\text{HP})
    - C_\text{LP}^\text{2pt}(t; t_0, \mathbf{x}_i^\text{HP})
    \right] \,,
  \label{eq:2pt_est}
\end{align}
where $C_\text{LP}^\text{2pt}$ and $C_\text{HP}^\text{2pt}$ are the two-point
function calculated in LP and HP, respectively, 
$\mathbf{x}_i^\text{LP}$ and $\mathbf{x}_i^\text{HP}$ are the different source
positions, and $N_\text{LP}$ ($N_\text{HP}$) is the number of source positions 
where the two-point function is evaluated in LP (HP).
%

In this study, we spread 60 LP and 4 HP source positions in four timeslices,
and calculate LP estimate of the two-point function by truncating the Dirac 
inversion with low-accuracy stopping criterion, $r_\text{LP}\sim 10^{-3}$.
Since we need to perform multiple Dirac inversions on a lattice, deflating the 
low-eigenmodes is an efficient way of reducing the computational cost.
Here we use multigrid solver \cite{Osborn:2010mb} to deflate the low-eigenmodes.

\subsection{Quark Loop}
The quark loops
$\sum_\mathbf{x} \Tr [M^{-1}(\tau, \mathbf{x}; \tau, \mathbf{x}) \Gamma]$
are estimated by a stochastic method.
Let us consider random noise vectors $\vert \eta_i \rangle$ in color/spin/spacetime 
space, satisfying
\begin{align}
 \frac{1}{N} \sum_{i=1}^N \vert \eta_i \rangle 
  = \mathcal{O}\left(\frac{1}{\sqrt{N}}\right) \,, \qquad
 \frac{1}{N} \sum_{i=1}^N \vert \eta_i \rangle \langle \eta_i \vert 
  = \mathbb{1} + \mathcal{O}\left(\frac{1}{\sqrt{N}}\right) \,.
\end{align}
Having the corresponding solutions $\vert s_i \rangle$ of the Dirac equation 
$M \vert s_i \rangle = \vert \eta_i \rangle$, one can stochastically estimate 
the inverse Dirac matrix as well as the quark loop by
\begin{align}
 M^{-1} 
  = \frac{1}{N} \sum_{i=1}^N \vert s_i \rangle \langle \eta_i \vert 
    + \mathcal{O}\left( \frac{1}{\sqrt{N}} \right)\,, \qquad
 \sum_\mathbf{x} \Tr [M^{-1}(\tau, \mathbf{x}; \tau, \mathbf{x}) \Gamma]
 \approx \frac{1}{N} \sum_{i=1}^N \langle \eta_i \vert_\tau \Gamma \vert s_i \rangle_\tau \,,
  \label{eq:stoch_eval}
\end{align}
where $\vert x \rangle_\tau$ is a vector whose $t=\tau$ components are filled by 
$\vert x \rangle$, and other components are zero.
%

For the stochastic estimation, we adopt the truncated solver method (TSM) 
\cite{Bali:2009hu} to reduce computation time.
Similar to the AMA, which we use in the two-point function calculation, most of 
the calculation is done in LP, and the result is corrected by small number of HP 
calculation:
\begin{align}
M_E^{-1} 
  = \frac{1}{N_\text{LP}} \sum_{i=1}^{N_\text{LP}} 
      \vert s_i \rangle_\text{LP} \langle \eta_i \vert
   + \frac{1}{N_\text{HP}} \sum_{i=N_\text{LP}+1}^{N_\text{LP} +N_\text{HP}} 
      \left( \vert s_i \rangle_\text{HP} 
        - \vert s_i \rangle_\text{LP}\right) \langle \eta_i \vert
\,.
\label{eq:stoch_m_inv}
\end{align}
Here $\vert s_i \rangle_\text{LP}$ and $\vert s_i \rangle_\text{HP}$ are the 
LP and HP solution of the Dirac equation for a given source $\vert \eta_i \rangle$.

For the random noise source, we tested various type of random numbers such as 
$\mathbb{Z}_2$, $\mathbb{Z}_2 \otimes i\mathbb{Z}_2$, $\mathbb{Z}_4$ and Gaussian.
It turns out that all of them are similar in terms of the total statistical error, but 
the Gaussian random number is marginally better in our case.
However, if one does not use the hopping parameter expansion, which will be discussed 
later in this section, $\mathbb{Z}_N$ type random noises will be a better choice as they
cancel part of leading noises.

The scaling form of the total statistical error of disconnected quark loop 
is \cite{Blum:2012uh}
\begin{align}
\label{eq:err_imp}
 \sigma^\text{imp} \approx
 \sigma \sqrt{\frac{1}{N_\text{LP}} + \frac{C}{N_\text{HP}}}
 = \sigma \sqrt{\frac{1}{N_\text{LP}} (1 +  R_{\text{LP}/\text{HP}} \times C )}\,,
\end{align}
where $R_{\text{LP}/\text{HP}} \equiv N_\text{LP}/N_\text{HP}$, and $C$ is 
determined by the difference (correlation) between $\vert s_i \rangle_\text{HP}$ 
and $\vert s_i \rangle_\text{LP}$.
In other words, the error scales as ${\sqrt{1/N_\text{LP}}}$, and the contribution 
from the correction term is determined by $R_{\text{LP}/\text{HP}}$ and accuracy
of the LP estimate.
In order to maximize computation efficiency, we use $R_{\text{LP}/\text{HP}}=30$, 
and calculate $\vert s_i \rangle_\text{LP}$ by the multigrid solver with
$r_\text{LP} \sim 5\times 10^{-3}$.

Required $N_\text{LP}$ is determined by the scaling of total error in disconnected
contribution to the nucleon charges, which is a combination of the error from the 
nucleon two-point function and that of the disconnected quark loop.
$N_\text{LP}$ dependence of the total error can be written as $\sigma = 
\sigma_\infty \sqrt{1 + X_\text{TSM}/N_\text{LP} }$. 
Here $\sigma_\infty$ is the error when $N_\text{LP} \rightarrow \infty$, so it is 
determined by the precision of the two-point function.
$X_\text{TSM}$ is determined by the gamma structure of the operator and TSM 
parameters.
On a12m310 lattice, with all the error reduction techniques given in this paper,
we find that $X_\text{TSM} \approx 40, 650$ and $15000$ for $g_S$, $g_A$ and $g_T$,
respectively, for light quarks.
In other words, $\mathcal{O}(100)$ of random noise sources is enough for the 
scalar, but $\mathcal{O}(1000) \sim \mathcal{O}(10000)$ is needed for the tensor.
In this study, we use $N_\text{LP} = 5000[1500], 11000[4000]$ and $4000[1200]$ on
a12m310, a12m220 and a09m310 lattices for light[strange] quarks.
%

For error reduction, we use the hopping parameter expansion (HPE) 
\cite{Thron:1997iy, Michael:1999rs}.
The Dirac matrix, written in the form of $M = \frac{1}{2\kappa}\left(\mathbb{1} - 
\kappa D \right)$ with a hopping parameter $\kappa$ can be expanded as
\begin{align}
 \frac{1}{2\kappa} M^{-1} 
  = \mathbb{1} + \sum_{i=1}^{k-1} (\kappa D)^i + (\kappa D)^k \frac{1}{2\kappa} M^{-1}
\,,
\end{align}
up to any order of $k$. 
Taking $k=2$, the disconnected quark loops are given as
\begin{align}
 \Tr\left[M^{-1} \Gamma \right] 
 = \Tr\left[\left(2\kappa \mathbb{1} + 2\kappa^2 D 
    + \kappa^2 D^2 M^{-1}\right) \Gamma \right]
\,.
\end{align}
Here the first term inside the trace of r.h.s can be easily calculated by hand, 
and the second term cannot contribute to the nucleon charges.
Hence what we need to evaluate is only the third term, 
$\Tr\left[\kappa^2 D^2 M^{-1} \Gamma \right]$.
Since noise terms are removed, HPE allows us to reduce the statistical error.

One of the major sources of the statistical noise in the stochastic estimation of
the disconnected quark loops is the correlation between the spacetime/spin/color
components.
Hence one may reduce the noise by (1) dividing those components into $m$ subspaces 
with proper separations, (2) evaluating the quark loops in those subspaces and (3)
obtaining the full results by summing the results of those subspaces.
This is the dilution technique \cite{Bernardson:1994at, Viehoff:1997wi, 
Foley:2005ac}.
In dilution, the computation cost increases as it needs to be evaluated on all $m$ 
subspaces.
As a results, it is useful only if the gain from the noise reduction is bigger 
than the increase of computation cost.
We do not find any error reduction from the dilution in spin or color components
on nucleon charges.
Dilution in time direction reduces error, but the gain is similar or smaller than
the increase of computation cost.

We average all the possible combinations of the gamma structures, and 
forward/backward propagations.
We also set zero for the real or imaginary components that should be zero by the
$\gamma_5$-hermiticity of clover Dirac operator, $M^\dag = \gamma_5 M \gamma_5$.

\section{Results and Conclusion}
\begin{figure*}[tb]
\centering{
  \subfigure{
    \includegraphics[width=0.7\linewidth]{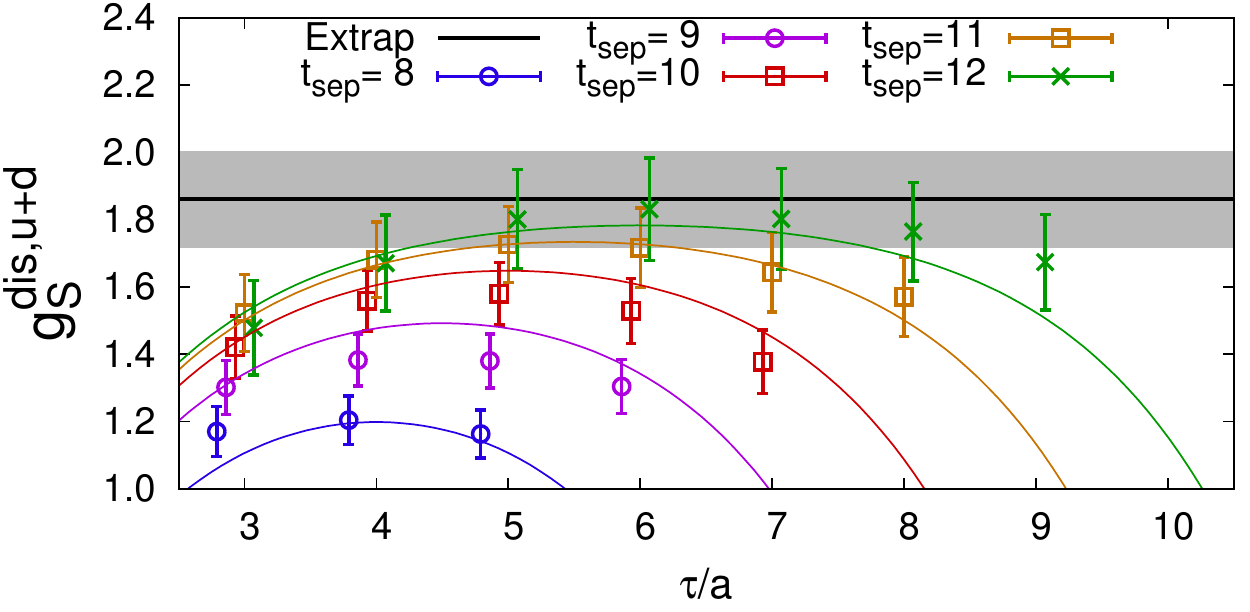}
  }\\
  \subfigure{
    \includegraphics[width=0.7\linewidth]{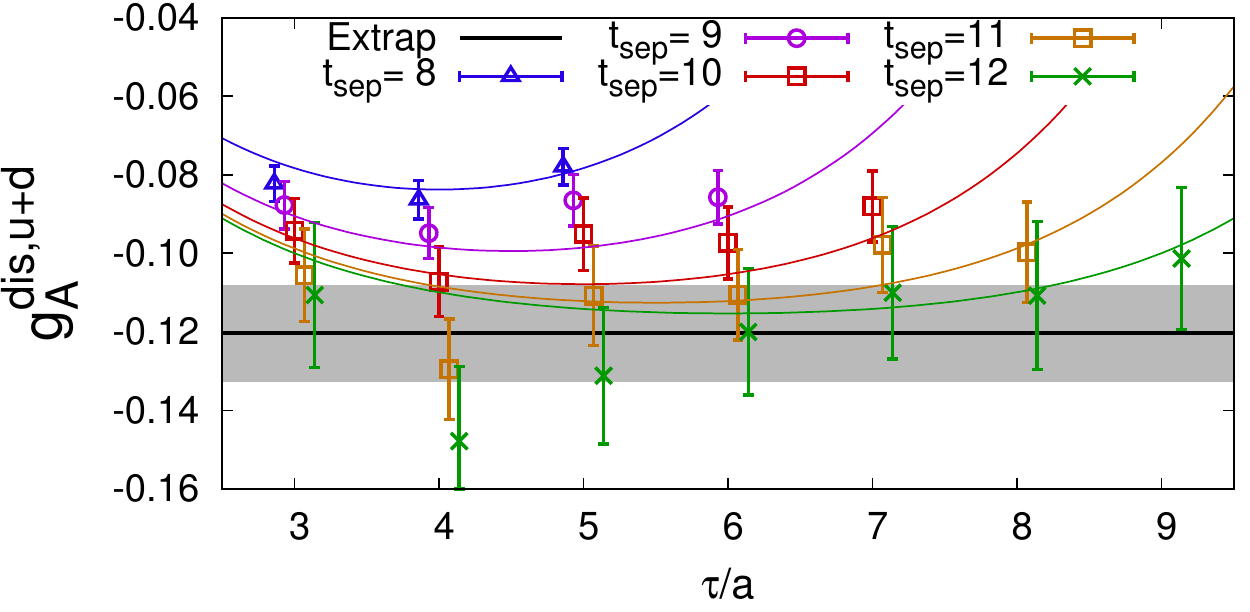}
  } \\
  \subfigure{
    \includegraphics[width=0.7\linewidth]{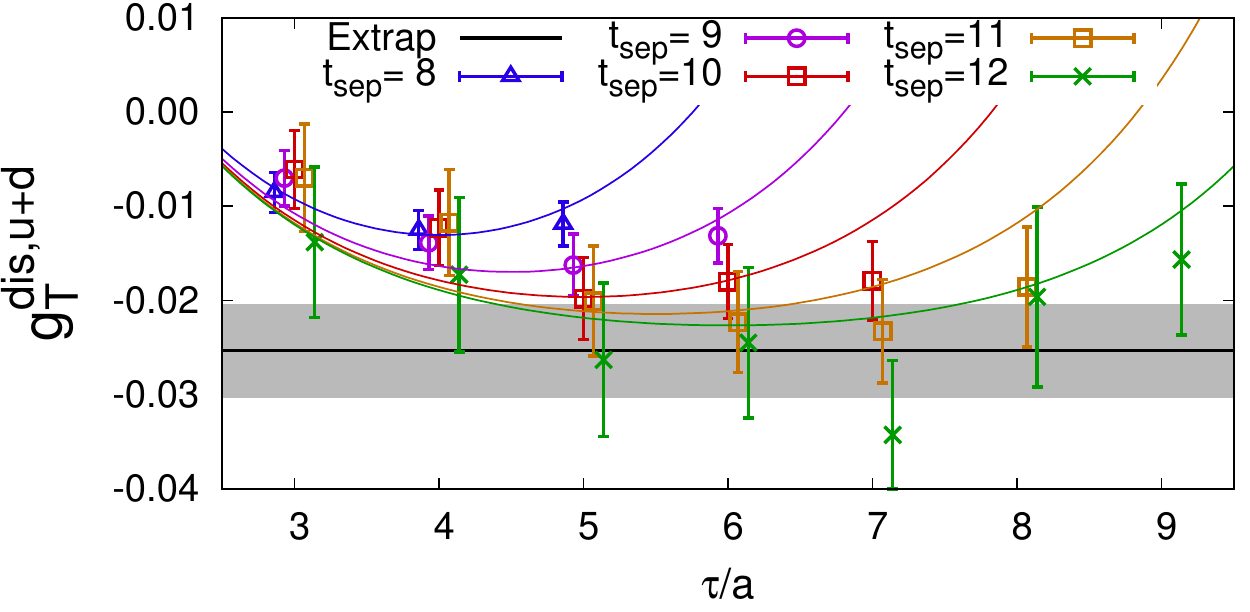}
  }
\caption{ Disconnected contribution to the nucleon charges for different
  source-sink separation $t_\text{sep}$ and operator insertion $\tau$ in
  lattice unit on a12m310 ensemble.
  Those are extrapolated to the limit of $t_\text{sep} \gg \tau \gg 0$, plotted as
  a black solid line with gray error band.
  \label{fig:res_a12m310}}
}
\end{figure*}

The nucleon charges should be calculated with the ground state nucleons.
On the lattice, however, nucleon state is generated by an interpolating 
operator, and it also generates the excited states of the nucleon.
We remove the contamination from the excited states by fitting the results to
a fit function including terms up to one excited state:
\begin{align}
C^\text{2pt}(t_f,t_i) = 
  &{|{\cal A}_0|}^2 e^{-M_0 (t_f-t_i)} + {|{\cal A}_1|}^2 e^{-M_1 (t_f-t_i)}\,,\\
C^\text{3pt}_{\Gamma}(t_f,\tau,t_i) = 
  &|{\cal A}_0|^2 \langle 0 | \mathcal{O}_\Gamma | 0 \rangle  e^{-M_0 (t_f-t_i)} +
   |{\cal A}_1|^2 \langle 1 | \mathcal{O}_\Gamma | 1 \rangle  e^{-M_1 (t_f-t_i)} +{}\nonumber\\
  &{\cal A}_0{\cal A}_1^* \langle 0 | \mathcal{O}_\Gamma | 1 \rangle  e^{-M_0 (\tau-t_i)} e^{-M_1 (t_f-\tau)} +{}
   {\cal A}_0^*{\cal A}_1 \langle 1 | \mathcal{O}_\Gamma | 0 \rangle  e^{-M_1 (\tau-t_i)} e^{-M_0 (t_f-\tau)} ,
\label{eq:2pt_3pt}
\end{align}
where $t_i=0$, $t_f = t$, and $|0\rangle$ and $|1\rangle$ are the ground and first 
excited nucleon states, respectively.
Details of the fitting procedure is described in Ref.~\cite{Bhattacharya:2013ehc}
as the {\it two-simRR} method.
Extrapolation plots on a12m310 ensemble are given in Fig.~\ref{fig:res_a12m310}.

The extrapolated results are renormalized at $2\GeV$ in $\overline{\text{MS}}$
scheme by using the non-perturbative renormalization (NPR) only with the connected 
diagrams.
As an approximation, disconnected diagrams and possible mixing with other quark 
flavors are neglected \cite{QCDSF:2011aa}.
The quark mass difference between light and strange quarks is also neglected; the
charges for the strange quarks are renormalized by those of light quarks.
The NPR procedure is described in Ref.~\cite{Bhattacharya:2013ehc}.
\begin{table}
\setlength\dashlinedash{1.5pt}
\setlength\dashlinegap{1.5pt}
\begin{center}
\begin{tabular}{c|rr|rr|rr}
\hline\hline
\multirow{2}{*}{Charge}    &  \multicolumn{2}{c|}{a12m310} & \multicolumn{2}{c|}{a12m220} & \multicolumn{2}{c}{a09m310} \\ \cdashline{2-7}
          & Conn. & Disconn. & Conn. & Disconn. & Conn. & Disconn. \\
\hline
$g_S^{u+d}$ &  5.22(23) &     1.65(14) &  6.16(33) &     1.49(36) &  6.01(25) &     1.40(14) \\
$g_A^{u+d}$ & 0.581(25) & $-$0.116(12) & 0.587(31) & $-$0.170(41) & 0.628(22) & $-$0.103(16) \\
$g_T^{u+d}$ & 0.619(26) & $-$0.024(05) & 0.625(31) & $-$0.005(11) & 0.622(25) & $-$0.011(05) \\ 
\hdashline
$g_S^{s}  $ &         0 &    0.584(48) &         0 &    0.639(81) &         0 &    0.440(44) \\
$g_A^{s}  $ &         0 & $-$0.036(05) &         0 & $-$0.038(10) &         0 & $-$0.022(07) \\
$g_T^{s}  $ &         0 & $-$0.0027(24)&         0 & $-$0.0009(32)&         0 &    0.0038(34)\\
\hline\hline
\end{tabular}
\caption{ Preliminary results of the connected and disconnected quark loop 
contributions to isoscalar nucleon charges of light quarks and nucleon charges
of strange quarks on three different ensembles.
Those are renormalized at $2\GeV$ in $\overline{\text{MS}}$ scheme, using
NPR only with the connected part.
}
\label{tab:res}
\end{center}
\end{table}

In Table~\ref{tab:res} we present our preliminary results of the connected and 
disconnected quark loop contributions to isoscalar nucleon charges of light quarks 
and nucleon charges of strange quarks on three different ensembles.
The calculation of the connected parts is described in 
Ref.~\cite{Bhattacharya:2013ehc}.
For disconnected $g_A^{u+d}$, we obtained slightly smaller values (about $2\sigma$) 
than those from authors of Ref.~\cite{Abdel-Rehim:2013wlz}.
Disconnected $g_T^{u+d}$ is yet to be compared as they have large statistical error.
For $g_A^{s}$, we obtained consistent value with Refs.~\cite{QCDSF:2011aa,
Abdel-Rehim:2013wlz,Engelhardt:2012gd} on a09m310 lattice, and slightly smaller value
on a12 lattices.
However, note that continuum and chiral extrapolations are needed to give a meaningful 
comparison.

In this study, we calculate the disconnected quark loop contribution to the 
nucleon charges.
Those are important observables for the neutron electric dipole moment (nEDM)
study through the quark EDM, and dark matter search.
The techniques of disconnected quark loops presented in this paper can be 
applied to the study of hadronic form factors, chromo EDM operators and 
transverse momentum dependent distribution function (TMD) observables 
\cite{Musch:2011er}.

\section*{Acknowledgments}
We thank the MILC Collaboration for sharing the HISQ lattices.
Simulations were performed using the Chroma software suite \cite{Edwards:2004sx} 
on facilities of the Institutional Computing at LANL, USQCD Collaboration funded 
by the U.S. DoE and Extreme Science, Engineering Discovery Environment (XSEDE) 
supported by NSF grant number OCI-1053575.
T.B., R.G. and B.Y. are supported in part by DOE Grant No. DE-KA-1401020.

\end{document}